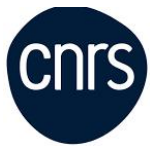 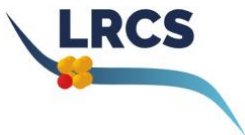 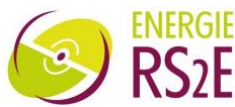 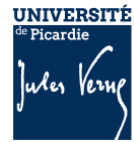

# Investigating Cathode Electrolyte Interphase Formation in NMC 811 Primary Particles Through Advanced 4D-STEM ACOM Analysis

Kevyn Gallegos-Moncayo[1,4], Justine Jean[1,2], Nicolas Folastre[1,2], Arash Jamali[1,3], Arnaud Demortière[1,2,4*]

[1.] Laboratoire de Réactivité et Chimie des Solides (LRCS), CNRS UMR 7314, 80009 Amiens, France.
[2.] Réseau sur le Stockage Electrochimique de l'Energie (RS2E), FR CNRS 3459, 80009 Amiens, France
[3.] Plateforme de microscopie, UPJV, Hub de l'Energie, 15 rue Baudelocque, 80039 Amiens, France.
[4.] ALISTORE-European Research Institute, CNRS FR 3104, 80039 Amiens Cedex, France

**\* Corresponding author:** arnaud.demortiere@cnrs.fr

**Abstract**

The study focuses on NMC811, a promising material for high-capacity batteries, and investigates the challenges associated with its use, specifically the formation of the Cathode Electrolyte Interphase (CEI) layer due to chemical reactions. This layer is a consequence of the position of the LUMO energy level of NMC811 that is close to the HOMO level of liquid electrolyte, resulting in electrolyte oxidation and cathode surface alterations during charging. A stable CEI layer can mitigate further degradation by reducing the interaction between the reactive cathode material and the electrolyte. Our research analyzed the CEI layer on NMC811 using advanced techniques such as 4D-STEM ACOM and STEM-EDX, focusing on the effects of different charging voltages (4.3 V and 4.5 V). The findings revealed varying degrees of degradation and the formation of a fluorine-rich layer on the secondary particles. Detailed analysis showed the composition of this layer differed based on the voltage: only LiF at 4.5 V and a combination of LiF and LiOH at 4.3 V. Despite LiF's known stability as a CEI protective layer, our observations indicate it does not effectively prevent degradation in NMC811. The study concludes that impurities and unwanted chemical reactions leading to suboptimal CEI formation are inevitable. Therefore, future efforts should focus on developing protective strategies for NMC811, such as the use of specific additives or coatings.

**Introduction**

Today, the necessity of battery materials has increased immensely due to the growing portable devices market and the energetic transition from fossil fuels to electric vehicles (EV)[1,2]. This exponential growth has pushed the scientific community to search new materials with an acceptable ration cost/ safety in Lithium-Ion batteries (LIBs) to be part of this energetic change. One of the battery materials that has been studied is layered cathode materials ($LiMO_2$, M transition metal) as NMC ($LiNiMnCoO_2$). NMC is based in an existing cathode material $LiCoO_2$ in which Co is partially replaced by Ni and Mn, obtaining as a result $LiNi_xMn_yCo_zO_2$ (x+y+z =1)[3]. In the case of NMC 811 the predominant transition metal (TM) is Ni, being the final chemical composition $LiNi_{0.8}Mn_{0.1}Co_{0.1}O_2$ (Ni-rich cathode material). Besides the energy density, the objective of this replacement is to reduce the quantity of Co because of its cost, safety and difficulties to recycle[4]. NMC 811 presents a high discharge capacity thanks to its Ni content, being the theoretical values ~200 mAh/g[5,6].

NMC cathode is usually used at a high voltage window (over 4.2 V) and presents capacity fading at a low cycle number due to different degradation mechanisms[7,8,9]. During the charge of batteries at high voltage, the phenomenon takes place at the interface of the liquid electrolyte and solid active material interface leading to the CEI formation. At a high stage of Li removal, chemical potential of the cathode material is shifted and approaches the HOMO energy of the electrolyte. This leads to oxidation of it and formation of a CEI layer which has an impact over Li transport as well as electronic configuration of TM sites in the cathode. The layer is composed of lithium carbonates and oxides, alkyl carbonates as result of solvent electrolyte oxidation parasitic reactions, and $Li_xPO_YF_Z$ oxidation products coming from the electrolyte salt. It has been show that for rich Ni cathode materials, it is important to take care of the LUMO/HOMO energies of electrode and the electrolyte, as well as nucleophilic affinity of the components[10,11,12,13].

In the study led by Iban Azcarate *et al.*[14], the reactivity of LP30 electrolyte (LiPF6 salt in a DiMethyl Carbonate/Ethyl Carbonate mixture), commonly utilized in LIBs was explored. Using both simple and double cell configurations with glassy carbon cathodes and Lithium metal anodes, the electrolyte's behavior was scrutinized through NMR and XPS techniques. The research revealed that at 4.2 V, Ethyl Carbonate (EC) was the initial electrolyte component to undergo oxidation. When the voltage increased to 4.8 V, Di-Methyl Carbonate (DMC) also began oxidizing, generating various derivative products. These products underwent further oxidation at 5.4 V. The study also observed depositions of inorganic species like LiF at 4.2 V, with the predominant deposition comprising mostly organic products, forming a non-passivating layer between 4.2 and 4.8 V. Above 4.8 V, the CEI layer predominantly consisted of inorganic compounds, enhancing passivation, although it remained unstable up to 5.4 V. Notably, the researchers achieved increased passivity and stability of the cell by maintaining it at a constant voltage for several hours.

In NMC cathode materials, capacity increases as a function of the amount of Ni. However, to reach the theoretical high capacity of Ni rich cathode materials, it is necessary to charge the battery up to voltage above 4.2 V leading to degradations of common LP30 electrolyte. In the work of *Noh et al.*[8] a quasi linear correlation between Ni content increasing and a decrease of safety and stability of the battery have been clearly demonstrated. This phenomenon can be attributed to changes in microstructure and chemical properties that correspond with alterations in Nickel content. In cathode materials with high Nickel content, it has been observed that during cycling in LP30 electrolyte, the instability of the CEI layer leads to a reaction between $Ni^{4+}$ and the electrolyte at an advanced de-lithiation stage. This reaction amplifies the $Li^+/Ni^{2+}$ cation mixing, thereby accelerating structural deterioration. Furthermore, structural decay is also exacerbated by changes in lattice parameters, particularly along the c-axis. These changes induce strains that cause ruptures in secondary particles, which then come into contact with the electrolyte, further contributing to the material's degradation [15,16,17,18].

CEI presence over NMC 811 particles has been reported in literature multiple times, as well as its importance for a correct performance of LIBs. Parasitic reactions and products reduce the available Li quantity in the cell, and due to the non-ionic conductivity of these products, impedes the utilization of the remaining lithium, ultimately leading to a decreased battery capacity.[19,20,21,22,23]. One of the strategies for reduce these reactions is the formation of an ionic conductive layer, stable during cycling and resistant to mechanical deformation. Some works focused in the introduction of additives to the electrolyte for the formation of stable CEI layer. In the work of Sen *et al.*[21] triallylamine (TAA) is proposed as an additive to eliminate the presence of parasite compounds as HF that cause cathode damage. In this work, the batteries with TAA additive electrolyte present better performances in capacity retention, in which the CEI layer is more uniform and compact, and there is no presence of cracks at the surface of NMC particles.

Another strategy founded in literature is NMC coating[13,24]. Looking deeply in the work of Bishnu P. *et al.*[13], it was proposed to applied electrochemical fluorination technique (ECF) to form a stable LiF layer at the surface of the particles. Pristine NMC811 forms a non-stable CEI layer composed by inorganic compounds as LiOH, $Li_2CO_3$ and $Li_2O$, these compounds are electric and ionic insulators, which result in the decreasing of capacity of the battery due to inaccessibility to Li ions on the particles. LiF layer is an ionic conductor as well as an electric insulator, avoiding the degradation of LP30 due to electrochemical reactions, but allowing the transportation of Li ions between the cathode and the electrolyte. The results show that the formation of a stable CEI at low voltages increase the cycling stability of NMC.

TEM methodologies have been extensively employed to conduct in-depth investigations of battery materials at various scales[25,26,27,28]. Notably, the 4D-STEM ACOM technique has been instrumental in achieving precise phase characterization within batteries, both *in situ* and post-mortem. This local analytical method offers a significant advantage in detecting compounds through electron diffraction patterns, as opposed to solely conducting elemental analysis (i.e., detecting individual species). Furthermore, 4D-STEM strikes an optimal balance between accessibility and resolution, especially when juxtaposed with other compound detection methods in battery research, such as Neutron Diffraction or X-ray Photoelectron Spectroscopy (XPS).[29,30,31,32,33]

In the study by Ankush *et al.* [34], the researchers investigated LMNO (Lithium Manganese Nickel Oxide) thin films *in situ* and post-mortem, utilizing a specially adapted liquid TEM sample holder for electrochemical analysis. This investigation revealed insightful details about the electrochemical behavior of LMNO, including the characteristic oxidation peaks of Nickel. Additionally, it highlighted the coexistence of amorphous and crystalline phases in LMNO and identified the formation of organic compounds resulting from electrolyte degradation. Despite the versatility of the 4D-STEM ACOM technique, it encounters limitations in liquid cell environments, primarily due to signal-to-noise reduction caused by multi-scattering effects from the liquid electrolyte's thickness. However, these challenges can be mitigated through applications such as ePattern suite software, as demonstrated in the work of Folastre *et al.*[35]. This algorithm employs registration and reconstruction methods to enhance the pattern identification and denoising of diffraction signals. Such advancements are pivotal in improving image quality and the signal-to-noise ratio, thereby enabling more reliable and accurate pattern analysis in TEM studies.

The objective of this research is to investigate the intrinsic characteristics and the genesis of the Cathode Electrolyte Interphase (CEI) in NMC811 coin cells through postmortem examination. This investigation employs an integrated approach utilizing SEM Energy Dispersive X-ray Spectroscopy (SEM-EDX) for the analysis of secondary particles, and STEM Energy Dispersive X-ray Spectroscopy (STEM-EDX) in conjunction with 4D-STEM ACOM for the examination of primary particles. This analysis is further correlated with the electrochemical performance observed during cycling.

**Results and discussion**

In this study, Nickel Manganese Cobalt (NMC) coin cells were electrochemically cycled against lithium metal within a potential range of 2.7 to 4.5 V. To investigate the impact of upper potential limits on the formation and efficiency of the cathode electrolyte interphase (CEI) layer, two distinct cutoff voltages were employed: a standard limit at 4.3 V (referred to as the 4.3V-limit) and an extended limit at 4.5 V (referred to as the 4.5V-limit). Over the course of 90 cycles, the development of the CEI layer was meticulously analyzed. This involved assessing its efficiency relative to the number of cycles and conducting a comparative analysis with the efficiencies of CEI layers documented in existing literature.

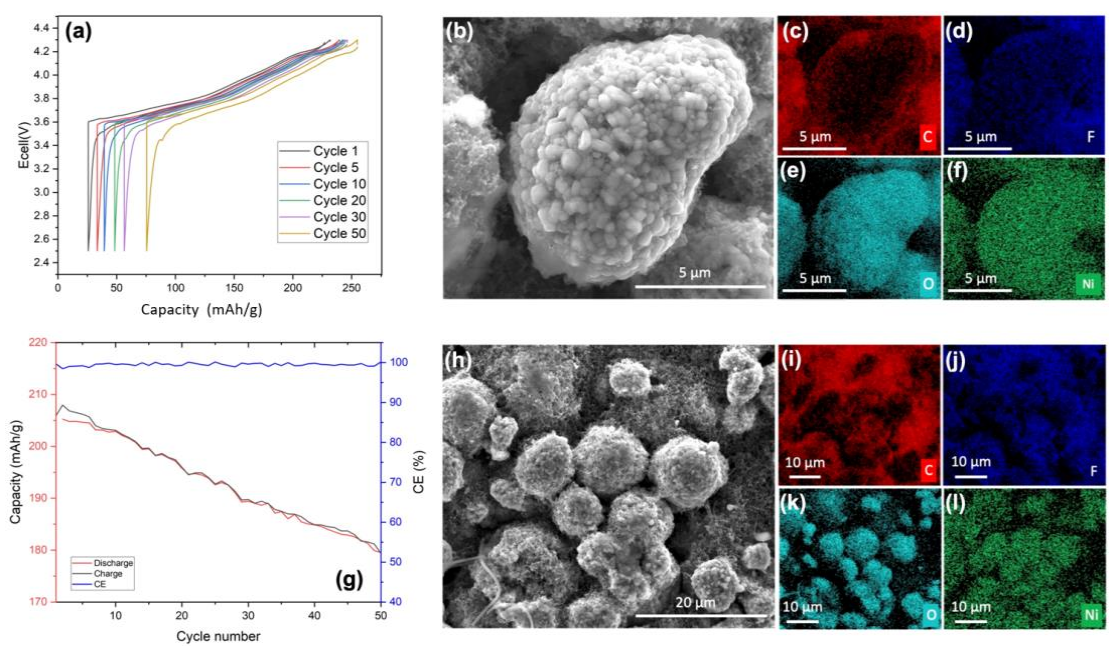

**Figure 1:** (a) Potential vs capacity electrochemical curve of cycled NMC at upper high voltage of 4.3 V, (b) secondary electron (SE) image of NMC single secondary particle in cathode sheet after cleaning (4.3V-limit cell). EDX analysis for elemental identification of (c) carbon, (d) fluorine, (e) oxygen, (f) nickel, (g) charge, discharge lost and CE in function of number of cycles on NMC811 coin cell, (h) secondary electron (SE) image of NMC secondary particles in cathode sheet after cleaning (4.3V-limit cell). EDX analysis for elemental identification of (i) carbon, (j) fluorine, (k) oxygen, (l) nickel.

In Figures 1a and 1g, we delineate the electrochemical characteristics of the sample. The cycling of the sample was conducted at a C-rate of C/20, interspersed with a 15-minute rest period between each charging and discharging cycle. Figure 1a depicts the potential versus capacity curve across various cycles. Up to the 20$^{th}$ cycle, minimal polarization is observed, indicative of the sample's stability and negligible degradation. However, at the 50$^{th}$ cycle, a marked increase in polarization is evident relative to previous cycles. The initial cycle demonstrates a capacity of approximately 200 mAh/g, which diminishes to around 143 mAh/g by the end of the battery's life cycle, culminating in a capacity retention of 71.5%. Figure 1g illustrates the progression of capacity loss and Coulombic efficiency (CE). A consistent pattern of degradation is observed. The absence of abrupt capacity loss at the graph's conclusion can be attributed to the fact that the span from 1 to 50 cycles represents merely a segment of the battery's potential state of life (SoL). This is corroborated by the CE data, which shows a modest decline of approximately 2% between the 1st and 50th cycles.

The polarization and the variations in charge/discharge efficiency, along with the coulombic efficiency (CE), can be primarily attributed to two distinct mechanisms leading to electrolyte degradation. These mechanisms include the formation of the CEI layer and the direct degradation of the electrolyte due to its interaction with lithium (Li) metal at the negative electrode. Li metal, especially, poses a significant risk due to its high reactivity when in direct contact with the LP30 electrolyte. This interaction results in the formation of various degradation products, such as 2,5-Dioxahexanedioic Acid Dimethyl, $CO_2$, CO, and phosphates, as referenced in studies.[36,37,38] It is also worth noting the presence of notable peaks between 3.4 V and 3.6 V during the analysis, likely induced by the periodic temperature fluctuations within the cycling room.

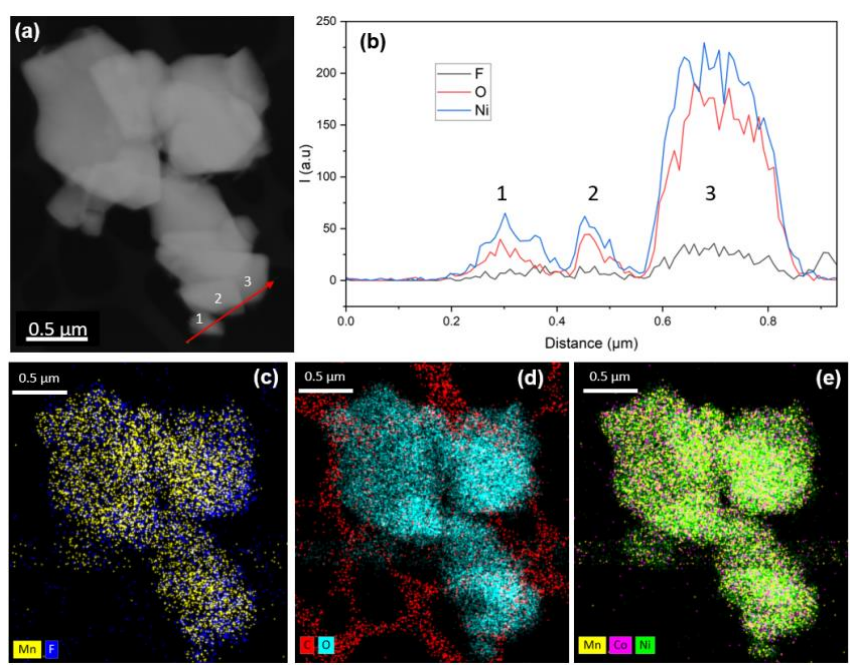

**Figure 2:** STEM-EDX analysis for primary agglomerate particles. (a) HAADF image, (b) profile spectra of selection area (red line) featuring Nickel, Oxygen and Fluorine presence over it, (c) cartography featuring Fluorine and Manganese over agglomerate for LiF observation, (d) cartography featuring Carbon and Oxygen for carbonates observation, (e) cartography featuring Nickel, Manganese and Cobalt (NMC).

The cathode material was examined using SEM, and the findings are exhibited in Figures 1c-f. Analysis of both the aggregated particles (Figures 1b-f, 1h-l) and individual particles (Figures 1d, 1j) revealed the presence of fluorine on the particle surfaces. Phosphorus was absent in the spectra, indicating that the detected fluorine originates from an electrochemical process rather than being a residual salt. This observation of fluorine suggests the formation of a CEI layer, which, according to existing literature, is likely to be a lithium fluoride (LiF) layer. In Figures 1c, 1d, 1i, and 1j, where both carbon and fluorine were observed, carbon distribution was non-uniform across the surface of the secondary particle. Conversely, fluorine exhibited a homogenous distribution, implying the absence of carbonate compounds in the CEI layer and suggesting the presence of lithium hydroxide (LiOH). However, the exact composition of the CEI layer remains undetermined due to analytical limitations. Additionally, in the Supplementary Information (Figure S2b), an elemental analysis including manganese is provided for comparison with fluorine, given their similar edge energies. The distinct presence of fluorine and manganese confirms that they are separate elements.

To gain deeper insight into the cell structure, analysis at the primary particle scale was conducted using STEM-EDX, with results presented in Figure 2. The High-Angle Annular Dark-Field (HAADF) imaging in Figure 2a reveals an agglomeration of primary particles, ranging in size from 0.5 to 1.5 µm. Figure 2b displays the spectral profile across various particle edges (indicated by a red line in Figure 2a), highlighting the presence of fluorine, particularly pronounced in the third particle examined. Figure 2c compares the Mn and F signals, showing a predominant Mn signal, with fluorine primarily detected along the edges of some particles, aligning with previous SEM-EDX observations. The C-O mapping in Figure 2d was conducted to assess the potential presence of carbonate compounds, but unlike the fluorine layer, no distinct carbon layer was observed. Finally, Figure 2e illustrates the transition metals, providing insights into the high homogeneity in composition of NMC particle with no segregation.

To enhance the comprehension of the results from STEM-EDX analysis, a comprehensive 4D-STEM investigation was conducted on two selected areas within the particle agglomerate. Crystallographic orientation and phase mapping were derived through the application of ACOM data processing methods[39,40], coupled with the utilization of the ePattern suite for data noise reduction[35]. Consistent with observations from the High-Angle Annular Dark-Field (HAADF) imaging, the agglomerate is identified as a conglomerate of primary particles. ACOM analysis was specifically applied to two zones, designated as zone 1 (illustrated in Figure 3a) and zone 2 (shown in Figure 3g), where STEM-EDX data indicated a high potential for CEI layer formation.

In zone 1, the crystal orientation and phase maps at a particle's edge are displayed in Figures 3b and 3d, respectively. Two primary components, NMC 811 and LiF, are identified, aligning with the STEM-EDX findings, as shown in Figures 3e and 3f. The orientation fidelity is notably high for the NMC phase, but substantially lower for LiF. Despite this, certain areas exhibit sufficient phase and orientation reliability, affirming the presence of LiF at the particle's edge. The diminished orientation reliability in some regions could be attributed to particle thickness or overlapping particle layers.

Regarding zone 2, the orientation mapping reveals a polycrystalline structure (Figure 3h). This finding is corroborated by TEM (Figure 3g) and STEM-HAADF (Figure 2a) imaging, which do not indicate particle superposition, thereby confirming the polycrystalline nature of the primary particles. This supports the theory of LiF layer instability due to mechanical disruption, as previously discussed. Figure 3l illustrates the uneven spatial distribution of LiF around the particle edge, mirroring observations in zone 1. Additionally, LiOH is detected in this area with high phase and orientation reliability (Figure 3m), yet it is notably absent from the particle edges.

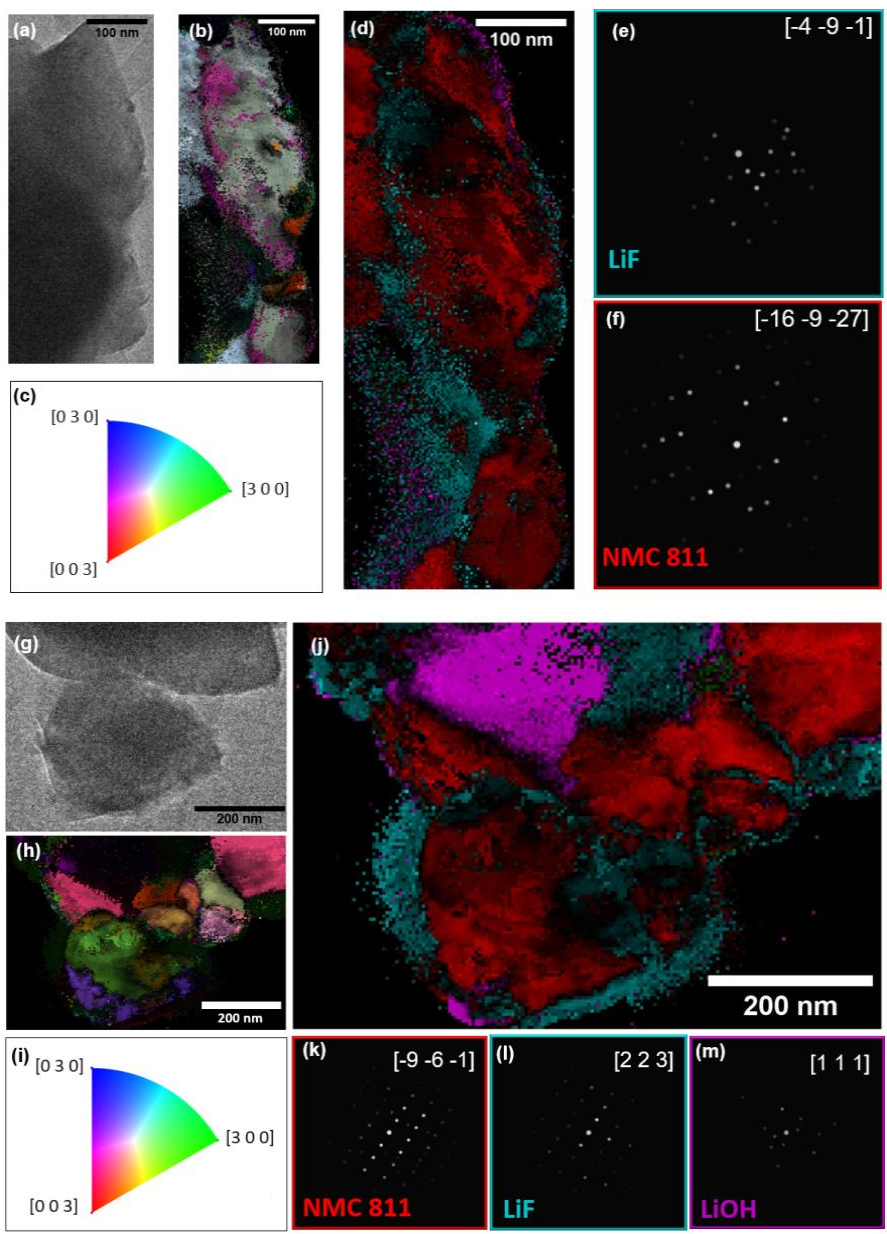

**Figure 3:** (a) TEM image of 4.3V-limit edge primary particle (zone 1). Particle edge 4D-STEM analysis: (b) reliability map and orientation map superposition, (c) orientation color map, (d) phase map and phase reliability map superposition, (e) LiF diffraction pattern, f) NMC 811 diffraction pattern, (g) TEM image of 4.3V-limit primary particle (zone 2). 4D-STEM primary particle analysis: (h) reliability map and orientation map superposition, (i) orientation color map, (j) phase map and phase reliability map superposition, (k) NMC 811 diffraction pattern, (l) LiF diffraction pattern, (m) LiOH diffraction pattern.

An alternative hypothesis for the origin of the LiF component layer and LiOH involves the potential presence of trace water within the cathode material, which could facilitate the production of HF and LiF[41,13]. Despite the samples being synthesized under meticulously controlled conditions within a dry room, the likelihood of moisture contamination during the battery assembly process cannot be discounted as a contributing factor to this observed phenomenon. Furthermore, it is plausible that residual water traces are inherently present in the electrolyte's solvents[42], which may predispose the formation of a LiF layer within the cathode-electrolyte interphase.

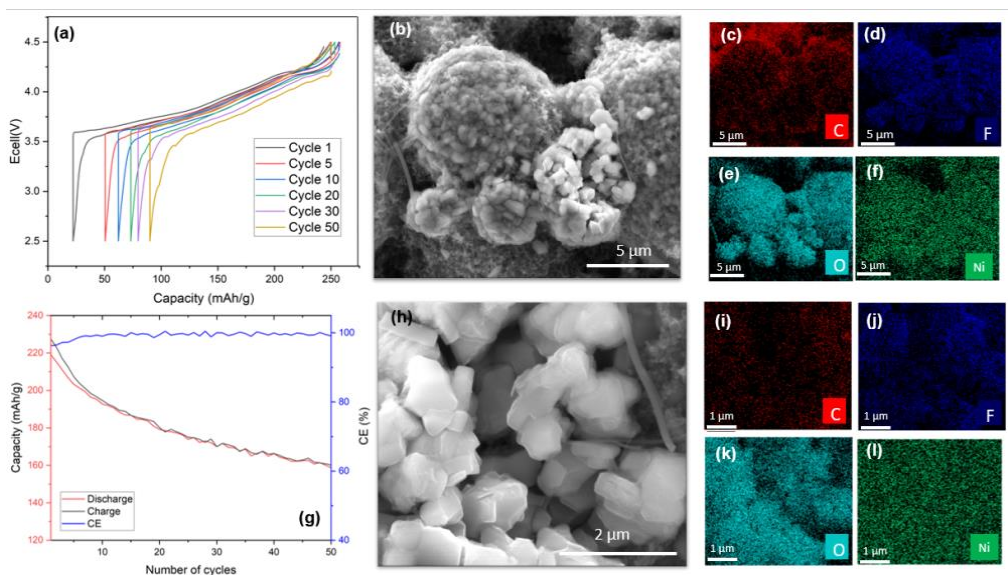

**Figure 4:** (a) Potential vs capacity electrochemical curve of cycled NMC at over the limit voltage of 4.5 V, (b) secondary electron (SE) image of NMC secondary particles in cathode sheet after cleaning (4.5V-limit cell). EDX analysis for elemental identification of (c) carbon, (d) fluorine, (e) oxygen, (f) nickel, (g) charge, discharge lost and CE in function of number of cycles on NMC 811 coin cell, (h) secondary electron (SE) image of NMC single secondary particles in cathode sheet after cleaning (4.5V-limit cell). EDX analysis for elemental identification of (i) carbon, (j) fluorine, (k) oxygen, (l) nickel.

In the presented study, electrochemical characteristics of a coin cell with a 4.5V voltage limit are elucidated, as depicted in Figure 4. This cell underwent cycling at a C/20 rate, incorporating a 15-minute interlude between successive charging and discharging phases. The polarization observed in the 4.5V-limited cell (Figure 4a) exhibits an enhanced magnitude compared to its 4.3V-limited counterpart, yet it maintains stability through the initial 50 cycles.

A closer examination of Figure 4b reveals a pronounced decline in capacity during the early stages of the cell's operational lifespan, stabilizing after approximately 20 cycles, similar to the pattern observed in the 4.3V-limited sample. This initial rapid degradation can likely be ascribed to the electrolyte's accelerated deterioration under high voltage conditions and the formation of a less stable SEI layer, which is susceptible to crystallographic and potential morphological alterations, in contrast to the 4.3V-limited scenario[43].

Furthermore, the Coulombic efficiency of the cell is illustrated in the same figure. While it remains relatively stable, there is an approximate 4% reduction, which is more significant than that of the 4.3V-limited sample. It is important to note that the thermal behavior of both cells was consistent, as they were subjected to identical cycling conditions in the same environment. For the 4.5V-limited cell, the initial capacity was approximately 230 mAh/g, diminishing to around 130 mAh/g at the end of its life cycle. This translates to a capacity retention rate of 56.5%, indicating a 21% decrease in efficiency compared to the cell cycled at a 4.3V cut-off voltage.

The primary aim of this study was to investigate the potential presence of CEI layer constituents, particularly carbonates, through SEM-EDX analysis. Figure 4b illustrates a cluster of secondary particles, with at least one exhibiting fracturing. As previously noted, mechanical strain during cycling may induce deformation, potentially leading to the fracturing of secondary particles, as introduced earlier in this manuscript. The detection of fluorine within the interior of the particle, as shown in Figure 4f, implies pre-cycling damage. This is consistent with observations from the sample at 4.3V-limit, where fluorine is present on the surface of all analyzed particles (Figures 4e-f), indicating the formation of a reactive and potentially unstable CEI layer during cycling. In contrast, the absence of carbon on the surface of particles in the 4.5V-limit sample suggests a lack of carbonate components. Supporting Information Figures S3c and S3d present the manganese cartography of the 4.5V-limit sample, demonstrating a distribution pattern distinct from that of fluorine, thus confirming the presence of the latter element and eliminating any potential misinterpretation of spectral energies.

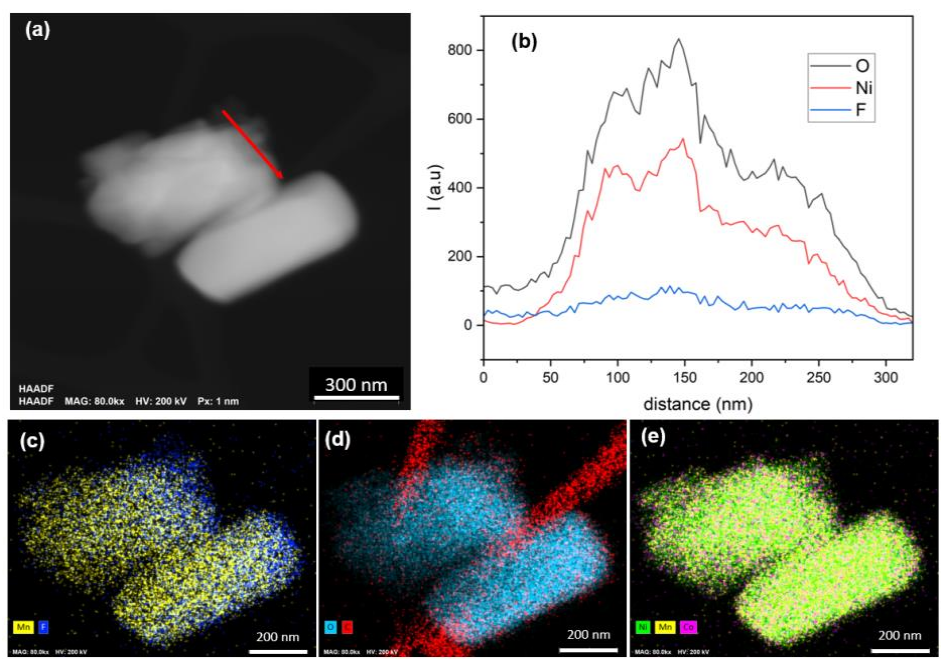

**Figure 5:** STEM-EDX analysis for primary particles a) HAADF image, b) profile spectra of selection area (red line figure 5a) featuring Nickel, Oxygen and Fluorine presence over it, c) cartography featuring Fluorine and Manganese over particles for LiF observation, d) cartography featuring Carbon and Oxygen for carbonates observation, e) cartography featuring Nickel, Manganese and Cobalt (NMC).

In the scientific analysis of the sample labeled 4.5V-limit, a thorough STEM-EDX examination of primary particles was conducted. The findings are presented in Figure 5. This analysis entailed a meticulous study of two primary particles, each approximately 500 nm in size. Figure 5c displays a detailed mapping of Manganese and Fluorine elements. The presence of Fluorine is evident along the edge of these particles. Consequently, a more targeted STEM-EDX profiling was carried out on one of these particles (as depicted in figures 5a and 5b), revealing an inhomogeneous Fluorine distribution along the edges. This inhomogeneity is attributed to the varying thickness of the particles.

Parallel to the procedure executed for the 4.3V-limit sample, a Carbon-Oxygen (C-O) mapping was performed. This step aimed to detect any potential carbonate layers. However, the Carbon signals detected around the particle edges were insufficient to conclusively identify a CEI layer composed of carbonate.

Furthermore, the comparative analysis of Figure 5e, showing NMC mapping, and Figure 5a, depicting STEM-HAADF imaging, provides a crucial insight. It confirms the absence of overlap between the analyzed particles. This observation is significant as it implies that 4D-STEM analysis could be effectively employed to ascertain the polycrystalline nature of these primary particles without any complications arising from particle superposition.

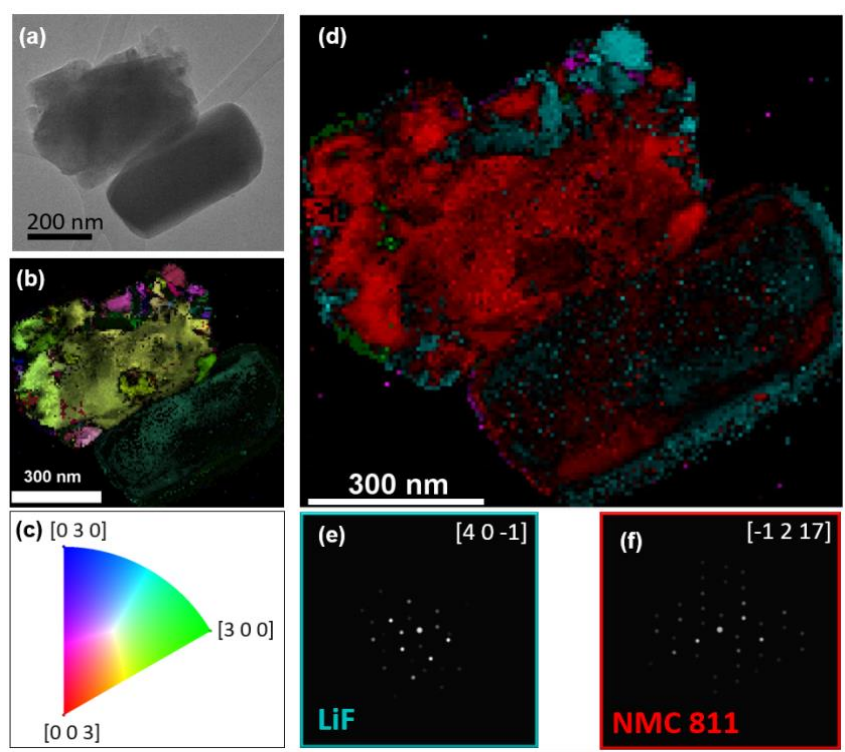

**Figure 6:** a) TEM image of 4.5V-limit primary particles, b) 4D-STEM analysis of primary particles: reliability map and orientation map superposition, c) orientation color map, d) 4D-STEM analysis of primary particles phase map and phase reliability map superposition, e) diffraction pattern (DF) for LIF (blue) component, f) DF for NMC component (red).

4D-STEM ACOM analysis was performed over the zone of interest, as shown in figure 6. First at all, the orientation map (figure 6b) shows a polycrystalline composition for the particle on the left of the scan. For the particle on the right, the thickness does not allow to have a high enough orientation/phase reliability to conclude on a polycrystalline or monocrystalline composition. This information allows us to reaffirm the hypothesis of lower performance of sample 4.5 V-limit in comparison to 4.3 V-limit due to layer rupture contribution to the deformation at higher voltage. In figure 6d, the phase map shows the presence of both NMC (figure 6f) and LiF (figure 6e), and the orientation and phase maps (together with STEM-EDX analysis) confirm the presence of LiF on particle edge, with higher confidence than the sample 4.3 V-limit. The higher confidence could be since a higher voltage, a more aggressive degradation of the electrolyte leads to more LiF formation. However, this does not mean a better protection from capacity loss as we could expect from the literature reviewed, in contrary, it rather represents a higher loss due to the non-stability of the CEI.

Considering the analysis of both samples, the LiF layer is not formed homogeneously over the entire particle edge. A possible explanation for this is the fact that the CEI formation occurs at the surface of secondary particles, meaning that during grinding, core particles do not present any CEI formation. Moreover, the CEI layer can be affected due to the mechanical energy induced by the grinding of the particles.

Li inorganic compounds were expected to be present in the CEI layer, as has been already reported in literature. Carbonates and oxides were completely absent, LiOH was found in sample cycled at 4.3 V in small quantity and LiF was present in both samples (4.3V and 4.5V) as the predominant component. Even though, LiF has been reported to be used as CEI protective layer against degradation in NMC as we presented in the introduction of this paper, in our case the samples still presented a retention capacity between 70% and 50%, these values are close to those founded in other works for NMC 811 samples with any treatment or additive against degradation.

**Conclusion**

Ni-rich cathode materials are known for the several degradation mechanisms present due to their high potential working condition. Chemical and mechanical degradation has been reported in numerous works. In this research, we explored the complex degradation mechanisms in Ni-rich cathode materials, known for their susceptibility to chemical and mechanical degradation under high-potential conditions. Our primary focus was to investigate strategies for mitigating these degradation processes without compromising the capacity of Ni-rich materials. A pivotal aspect of this endeavor involved the formation of a stable CEI layer.

Utilizing advanced characterization techniques such as 4D-STEM with automated crystal orientation mapping, along with electrochemical analysis, SEM-EDX, and STEM-EDX, we conducted an in-depth study of CEI layer formation in conventional NMC 811 coin cell configurations. These configurations included a lithium metal anode and LP30 electrolyte, examined at both limit cutoff voltage and beyond.

Our findings revealed the formation of a LiF-based CEI layer in both samples, potentially resulting from water traces in the cathode during preparation. While LiF is frequently lauded for its potential in mitigating degradation phenomena in battery systems, our observations indicated that this layer does not form prior to cycling, and its protective efficacy is not consistent with results reported by other research groups. Our electrochemical data further highlighted a more pronounced degradation at higher cutoff voltages (4.5V-limit), which could be attributed to the continuous formation and dissolution of the CEI layer. This process, coupled with particle deformation, exacerbates the degradation of the electrolyte at elevated voltages.

Our methodological approach and experimental setup proved effective in studying and understanding CEI formation. However, it is critical to note that CEI formation is predominantly a surface phenomenon at the secondary particle scale. This implies a reduced likelihood of observing primary particles with a CEI layer in TEM analysis, especially considering that sample preparation involves powder grinding, which may induce post-mortem degradation of the existing CEI layer.

In perspective, for a better comprehensive analysis of CEI formation, liquid electrochemical *in situ* TEM cycling analysis techniques, including STEM-EDX and 4D-STEM, are indispensable. Our study also highlights the inevitability of parasitic reactions, even in controlled environments, across both homemade and commercial batteries. These insights suggest that future advancements in battery performance may hinge on integrating electrolyte additives or adopting cathode material coatings, as explored in previous studies. This direction holds promise for enhancing the durability and efficiency of Ni-rich cathode materials in high-potential battery applications. Finally, it is imperative to conduct a more advanced examination of phase transformations occurring at the surface of Ni-rich NMC in relation to the applied electrochemical potential and the composition of the CEI layer. This investigation is crucial for advancing our understanding of the interplay between CEI layer formation and the alteration in crystallinity of NMC materials.

**Materials and methods**

4D-STEM ACOM analysis:

The TEM investigations were conducted using an accelerating voltage set to 200 kV. During the diffraction experiments, the camera length was meticulously maintained at 300 mm. A precession angle of 0.7° was employed, aimed at minimizing dynamical scattering effects. The condenser aperture was precisely configured to 10 µm, resulting in a convergence semi-angle of 0.4 mrad. For electron beam control, Gun lens 3 was utilized, with the spot size adjusted to 5. The electron dosage for 4D-STEM analysis was established at 150 e/Å²/s. Data processing of the 4D-STEM dataset was executed utilizing the ePattern Suite software, which facilitated denoising operations with a prominence value set at 5. Subsequently, the ASTAR software package (developed by Nanomegas) was applied for the reconstruction of phase and orientation maps. This was achieved through the Automated Crystal Orientation Mapping (ACOM) technique, which relies on a pattern matching algorithm. The resolution of each diffraction pattern was configured to 512x512 pixels. Acquisition of these diffraction patterns was carried out using a OneView CMOS camera, manufactured by Gatan, CA, USA.

TEM/STEM analysis:

In the context of TEM analysis, the powder from the cathode sheet was detached from the current collector, manually pulverized, and subsequently stored in an aerated glass flask. HAADF-STEM imaging was conducted using a Tecnai G2 F20 S-Twin (ThermoFisher) system, operated at an accelerating voltage of 200 kV and equipped with a C2 aperture of 70 µm. Additionally, for the acquisition of elemental maps, an energy-dispersive X-ray spectroscope (EDX, Xflash, Bruker, Germany) was employed in STEM mode.

SEM analysis:

An environmental SEM (ESEM), specifically the FEI Quanta 200 Field Emission Gun (FEG) model, augmented with an advanced energy-dispersive X-ray (EDX) microanalyzer (X-Max 80, Oxford Instruments Co., UK), was employed for detailed microstructural analysis. Imaging modalities including secondary electron (SE) and backscattered electron (BSE) techniques were utilized, conducted under a controlled high vacuum environment at an electron acceleration voltage range of 10 to 15 keV. EDX spectroscopy was consistently executed at an acceleration voltage of 15 keV to ensure optimal elemental characterization. For SEM imaging, the cathode sheet was subjected to examination in its post-recovery state, following a meticulous cleaning process within a controlled atmosphere glovebox.

Electrochemical analysis:

The positive electrode material, namely NMC 811, was procured from NEI Corporation in the form of cathode sheets, each measuring 127 mm by 254 mm, and adhered to an aluminum current collector with a thickness of 16 µm. The composition of the cathode involved a blend of NMC 811 (constituting 90% of the active material), polyvinylidene fluoride (PVDF) as a binder (5%), and Super P conductive carbon (5%), achieving a uniform thickness of approximately 60 µm ± 5%. For the electrolyte, 100 µl of LP30, a commonly utilized formulation comprising lithium hexafluorophosphate (LiPF6) in a binary solvent system of ethylene carbonate (EC) and dimethyl carbonate (DMC) in a 1:1 volumetric ratio, was employed. A 1 mm thick fiberglass separator was integrated into the design. Lithium metal, serving as the negative electrode, was sourced from Sigma Aldrich, preserved in a controlled dry environment, and prepared as a thin foil for subsequent processing.

The coin cell assembly was conducted within a dry room environment, utilizing a 13 mm punched cathode, a 15 mm punched separator, and an 8 mm punched lithium metal. Post-cycling, the coin cell was disassembled within a nitrogen-filled glovebox, and the cathode sheet was subjected to a thorough cleaning process using DMC. This cleaning involved a three-cycle spinning protocol, each at 3600 rpm for 3 minutes, followed by a drying phase at 80°C for one hour in an air atmosphere. Electrochemical characterization was performed by cycling the coin cell using a BSC-COM Biologic potentiostat, with the operational parameters and data acquisition managed by EC-Lab software, version 11.34.

**Acknowledgments**

As a part of the DESTINY PhD program, this publication is acknowledged by funding from the European Union's Horizon2020 research and innovation program under the Marie Skłodowska-Curie Actions COFUND (Grant Agreement #945357)". A part of the funding has been provided by the French Research Agency (ANR) as part of the DestiNa-ion Operando project (ANR-19-CE42-0014).

**Conflicts of interest**

The author report no conflicts of interest

**Supporting information**

Supporting information shows SEM-EDX images of secondary particles featuring Mn and F. The reference of the cif files used in 4D STEM and more details of the acquisition conditions are presented as well.

# Supporting Information

## Investigating Cathode Electrolyte Interphase Formation in NMC 811 Primary Particles Through Advanced 4D-STEM ACOM Analysis

Kevyn Gallegos-Moncayo[1,4], Justine Jean[1,2], Nicolas Folastre[1,2], Arash Jamali[1,3], Arnaud Demortière[1,2,4*]

[1.] Laboratoire de Réactivité et Chimie des Solides (LRCS), CNRS UMR 7314, 80009 Amiens, France.
[2.] Réseau sur le Stockage Electrochimique de l'Energie (RS2E), FR CNRS 3459, 80009 Amiens, France
[3.] Plateforme de microscopie, UPJV, Hub de l'Energie, 15 rue Baudelocque, 80039 Amiens, France.
[4.] ALISTORE-European Research Institute, CNRS FR 3104, 80039 Amiens Cedex, France

\* **Corresponding author:** arnaud.demortiere@cnrs.fr

**4D-STEM analysis**

<u>General parameters</u>

Library for bank generation in ASTAR was obtained using CIF file from ICSD website, the files used for all the searched programs were:

-NMC 811 collection code 143110

-LiOH collection code 27543

-$Li_2CO_3$ collection code 66942

-LiF collection code 41409

<u>Sample (4.3 cut off voltage)</u>

For sample 4.3V-limit (figure 3b-e in the article) the following analysis condition were used:

Lacey carbon films 200 MESH grind was used for particle analysis. The scanning conditions for particle of zone 1 where 110 pixels X, 270 pixels Y, step width of 15 giving as a result a scanning area of 0.41 by 1.01 µm and a step size of 3.75 by 3.75 nm. The exposition time for each individual scan (diffraction pattern) was 20 ms

For zone 2, the scanning conditions were 190 pixels X, 140 pixels Y, step width of 15, giving as result a scanning area of 0.71 by 0.53 µm and a step size of 3.75 by 3.75 nm. The exposition time for each individual scan (diffraction pattern) was 20 ms

Sample (4.5 cut off voltage)

For sample 4.5V-limit (figure 6a-e in the article) the following analysis condition were used:

Lacey carbon films 400 MESH grind was used for particle analysis. The scanning conditions for particles where 150 pixels X, 150 pixels Y, step width of 15 giving as a result a scanning area of 1.13 by 1.13 µm and a step size of 7.5 by 7.5 nm. The exposition time for each individual scan (diffraction pattern) was 20 ms

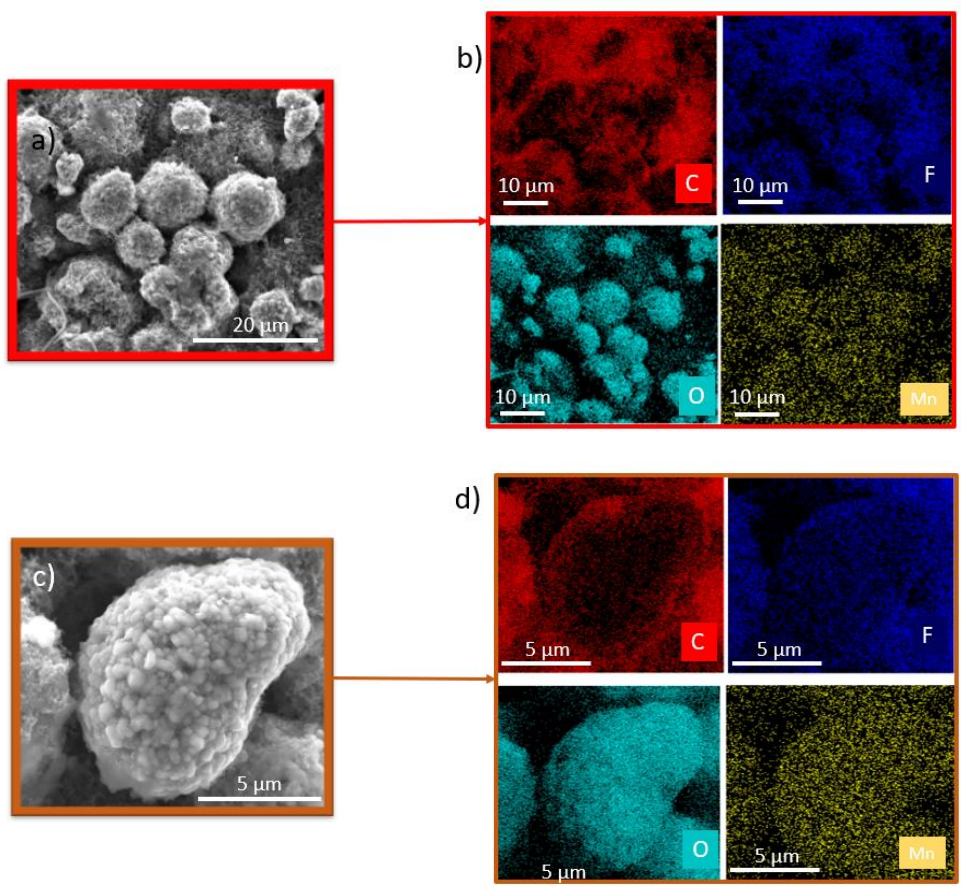

Figure S2: (a) secondary electron (SE) image of NMC secondary particles in cathode sheet after cleaning from 4.3V-limit cell, (b) EDX analysis for elemental identification (c) secondary electron (SE) image of NMC single secondary particle in cathode sheet after cleaning from 4.3V-limit cell, (d) EDX analysis for elemental identification.

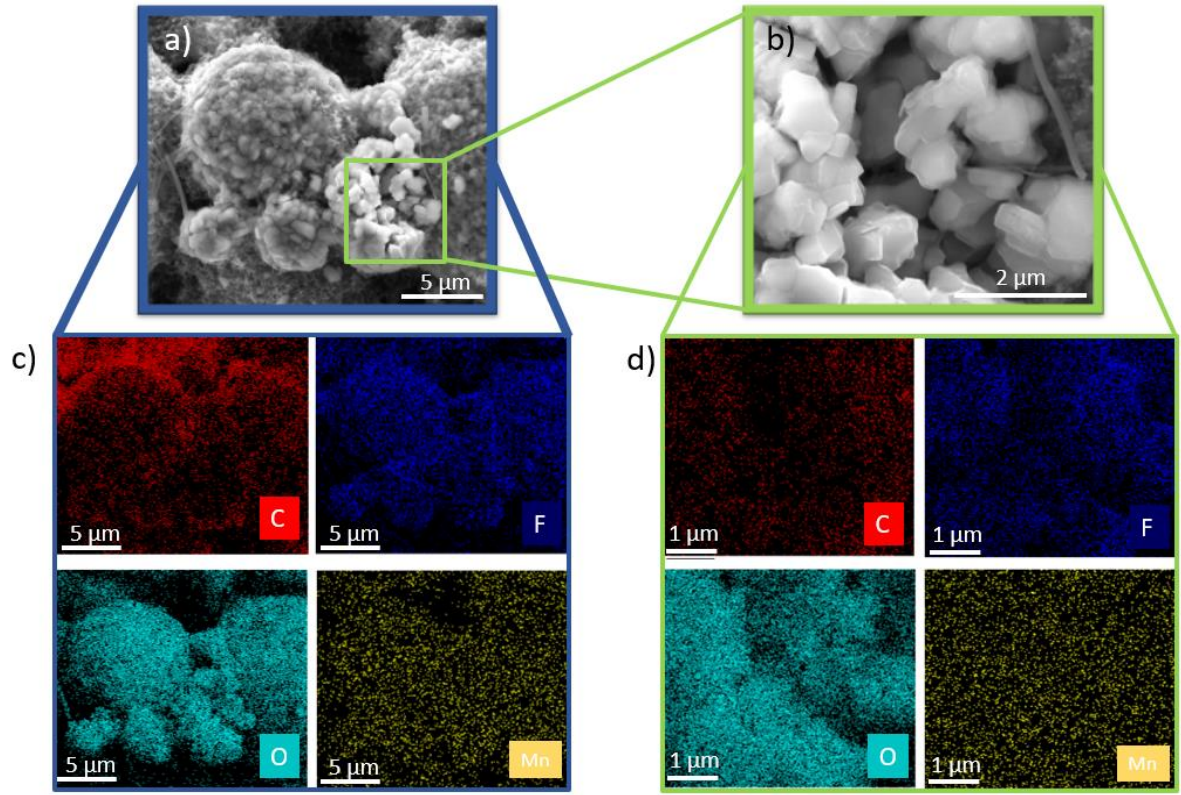

Figure S3: (a) secondary electron (SE) image of NMC secondary particles in cathode sheet after cleaning from 4.5V-limit cell, (b) zoom in over cracked particle in (a), (c) EDX analysis for elemental identification of different particles (d) EDX analysis for elemental identification in cracked particle.

The edge energy for Mn and F are similar, this could give as result the misinterpretation of F as Mn in the analysis software, for this reason both, Mn and F cartography, are presented for the sample 4.3V-limit (figure S2) and 4.5V-limit (figure S3). In both cases for the secondary particle agglomerate (figure S2a-b and figure S3a-c) and single secondary particle (figure S2c-d and figure S3b-d) the distribution of Mn in regards of F is not equal, allowing us to affirm the presence of both elements.